\documentclass{cjaa}                   % preprint, the final version for publication
\usepackage{graphicx}                   %for PS/EPS graphics inclusion, new
\input{epsf.sty}                        %for PS/EPS graphics inclusion, old
\input{psfig.sty}                       %for PS/EPS graphics inclusion, old

\begin{document}

   \title{Magnetic fields in our Galaxy: How much do we know?
%\footnotetext{$*$ Supported by the National Natural Science Foundation of China.}
}
%   \subtitle{Progress in the last decade}
   \subtitle{III. Progress in the last decade} 
%   *** I don't understand the III.

   \volnopage{Vol.0 (200x) No.0, 000--000}      %%preserved for Editor. DOn't remove!
   \setcounter{page}{1}          %%starting page, preserved for Editor. DOn't remove!

   \author{J. L. Han
     % \inst{}
      \mailto{hjl@bao.ac.cn}
%% Please move "\mailto{}" to the corresponding author of the paper.
%   \and R. N. Manchester
%      \inst{2}
      }
   \offprints{J. L. Han}                   %% is disabled in fact

   \institute{National Astronomical Observatories, Chinese Academy of Sciences,
             Beijing 100012, China\\
             \email{hjl@bao.ac.cn}
%% offprint requests will be sent. Note to pair \mailto{} with \email{}
%        \and
%             Australia Telescope National Facility, CSIRO, PO Box 76,
%        Epping, NSW 2121, Australia
          }

\vspace{-3mm}
\date{Received~~2006 month day; accepted~~2006~~month day}

\vspace{-5mm}
\abstract{ A decade ago, there was very limited knowledge of magnetic fields
of our Galaxy. The local fields in the Solar vicinity were known to be directed
towards a Galactic longitude $l\sim~90^{\circ}$ with reversed directions at
smaller Galacto-radii. The regular field strength was found to be about
2~$\mu$G.  The filaments near the Galactic Center show the possible poloidal
fields there. There was no information about the magnetic fields in the
Galactic halo. In last decade, there has been significant progress on
measurements of the Galactic magnetic fields.  In the Galactic disk, from
the RMs of a large number of newly observed pulsars, large-scale magnetic
fields along the spiral arms have been delineated in a much larger region
then ever before, with alternating directions in the arm and interarm
regions. The toroidal fields in the Galactic halo were revealed to have
opposite directions below and above the Galactic plane, which is an
indication of an A0 mode dynamo operating in the halo.  The strength of large-scale
fields also has been found from pulsar RM data to exponentially increase at
smaller Galacto-radii. Compared to the steep Kolmogorov spectrum of magnetic
energy at small scales, the large-scale magnetic fields show a shallow
broken spatial magnetic energy spectrum.
\vspace{-2mm}
          }

\authorrunning{J.~L. Han}            %author_head in even pages
\titlerunning{Magnetic fields in our Galaxy: How much do we know? III.}  % title_head in odd pages

   \maketitle
%% The author head (on even pages) and the title head (on odd pages) will be
%% automatically extracted from \author{} and \title{}. Whenever the title is too long,
%% you will be asked to supply a shorter one by inserting either \authorrunning{} or
%% \titlerunning{} before \maketitle. Anyway, you can specify your own heads in advance.
%%
%%
%% Note: In the following text body of your manuscript, please note several differences from
%%       other major journals:
%% (1) \subsection{Please Capitalize the First Letter of Each Notional Word in Subsection Title}
%% (2) Please Capitalize the First Letter of Each Notional Word in table's caption

%
%________________________________________________ sections below
%
%\vspace{-2mm}
\section{Introduction}           %% first-level sections will be auto-capitalized
\label{sect:intro}
%\hspace{15pt}%                   %% preserved for Editor

%% ChJAA editors DID NOT use \cite{} for citation, \ref and \label for
%% cross-references of Table/Figure in publication version.
%% ChJAA editors prefered you giving a citation as 'Michel et al. 1992', and
%% writting Table~1 or Fig.~1 and so forth. However, that will make authors
%% inconvenient in adjusting/adding/removing text, tables or figures. Anyway,
%% authors can use \cite, \citep and \citet as widely used in other journals.
%% ChJAA editors are moving to use a more flexible LaTeX source.

Magnetic fields in our Galaxy: How much do we know? When I first asked
myself this question, I discussed the magnetic fields in local Galactic disk
in Han (2001). The magnetic fields in the Galactic halo and the global field
structure picture with maximum likelihood were discussed in Han (2002) as
being the second part of the answer. Now, in this third one, I will talk 
about progress in the last decade, and also show how much we should know.

Only if the magnetic field $\vec{B}(x,y,z)$ in all positions in our Galaxy is
known, one can say that we have a complete picture of the Galactic
magnetic fields. Here, $x,y$ are in the Galactic plane, and $z$ is normal to that. In
practice, we only have ``partial'' measurements in some regions, so we 
never get a complete picture. However, if we ``connect'' the available
measurements of different locations, then we may outline some basic features
of the Galactic magnetic fields. 

Magnetic fields exist on all scales in our Galaxy, the Milky Way. When one
tries to look at the large-scale field structure, the small-scale fields act as
``random'' fields, ``interfering'' with your measurements to an extent that
depends on the strength of random fields. The large-scale magnetic fields
appear as a kind of smooth background at small scales, and exist in a
coherent manner at different locations inside our Galaxy. In this review,
the magnetic fields in our Galaxy we talk about are the fields in the
diffuse interstellar medium, rather than the fields in molecular clouds 
which are very extensively reviewed by Heiles \& Crutcher (2005).
%Many physical processes in interstellar medium modify the field
%structure. The supernova explosion compress the interstellar gas with
%frozen-in magnetic fields, stirring the large-scale fields to have more
%structures. The density wave can compress the random field to be anisotropic
%in a systematic pattern with respect to the spiral structure. The dynamo
%action can maintain or amplify the magnetic fields.

To describe the Galactic magnetic fields, we need to clarify following 
items:
\vspace{-1mm}
\begin{itemize}
\item {\bf Field structure}
  \begin{itemize}
      \item Disk field: local structure in the Solar vicinity (3 kpc)?
      \item Disk field: large scale structure and reversed directions
 in arm and interarm regions?
      \item Field structure in the Galactic halo?
      \item Field structure near the Galactic Center?
  \end{itemize}
\item {\bf Field strength $B$}
  \begin{itemize}
      \item Random field versus ordered field: $\langle \delta B \rangle^2$
      vs. $B^2$?
      \item Variation of field strength with the Galacto-radius
	($R=\sqrt{x^2+y^2}$), i.e. $B$ or $\delta B$ varies with $R$?
      \item Variation of field strength with the Galactic height ($z$): 
        $B$ or $\delta B$ varies with $z$?
      \item $B$ or $\delta B$: difference in arm and interarm regions?
  \end{itemize}
\item {\bf Fluctuations and scales}
  \begin{itemize}
      \item Spatial B-energy spectrum in large and small scales?
      \item Maximum field strength in the energy injection scale?
  \end{itemize}
\end{itemize}
\vspace{-1mm}
There are five observational tracers for the Galactic magnetic fields: 
Zeeman splitting, polarized thermal emission from the dusts in
clouds, polarization of starlight, synchrotron radio emission, Faraday
rotation of polarized sources. All knowledge of the Galactic magnetic
fields comes from observations of these tracers, but some show the local
fields and some show the large-scale fields. Compared to the knowledge a
decade ago about the above terms, significant progress has been made on
many aspects as we describe below.

\section{The Galactic magnetic fields: Status a decade ago}

\noindent{\bf A. Local disk field: there was some consensus.} 

Starlight polarization data are mostly limited to stars within 2 or 3 kpc
but gave the very first evidence for large-scale magnetic fields in
our Galaxy. It has been shown that the local field is parallel to
the Galactic plane and follows the local spiral arms (e.g. Andreasyan \&
Makarov 1989). The rotation measures (RMs) of small sample of local pulsars
showed that the local magnetic field going toward $l\sim90^o$
(Manchester 1974), with a strength about 2~$\mu$G. The field reversal near
the Carina-Sagittarius arm was shown by model-fitting to the
pulsar RM data by Thomson \& Nelson (1980). All these have been confirmed
later by much more pulsar RM data.

\vspace{1mm}
\noindent{\bf B. Which model for the large-scale disk field? Not clear.}

When available measurements are very limited, a good model is needed to
connect the measurements and give an idea of the basic features.
Simard-Normandin \& Kronberg (1980, SK80) and Sofue \& Fujimoto (1983) show
that the large-scale magnetic fields in the Galactic disk with a bisymmetric
spiral (BSS) structure of a negative pitch angle and field reversals at
smaller Galacto-radii can fit the (average) RM distribution of extragalactic
radio sources along the Galactic longitudes better than the concentric ring
model and the axisymmetric spiral (ASS) field model. After Hamilton \& Lyne
(1987) published the RMs of 185 pulsars, the field reversals near the
Carina-Sagittarius arm and the Perseus arm were confirmed by Lyne \& Smith
(1989). Rand \& Kulkarni (1989: RK89) failed to fit the BSS to the pulsar RM
data and emphasized the validity of the concentric ring model (also in Rand
\& Lyne 1994). Vall\'ee (1996) argued for an axisymmetric spiral field model
according to early RM data of extragalactic radio sources near tangential
directions of spiral arms. Han \& Qiao (1994: HQ94) carefully checked the
model and data, and found that the BSS model is the best to fit pulsar RM
data.

\vspace{1mm}
\noindent{\bf C. Fields in halo or thick disk? Did not know much.}  

SK80 and HQ94 found evidence for a thick magneto-ionic disk with a scale
height of $\sim$1.2~kpc. A thin and thick radio disk was found from modeling
the radio structure of our Galaxy at 408~MHz (Beuermann et al. 1985).
Large-scale polarized features in sky (see a comprehensive review by Reich
2006 and references therein) as well as the outstanding features in the RM
sky (e.g. SK80) were all attributed to the disturbed local magnetic fields.
No large-scale magnetic fields in the Galactic halo were recognized.

\vspace{1mm}
\noindent{\bf D. Fields near the Galactic center? Yes.} 

Near the Galactic center vertical filaments were observed (Yusef-Zadeh
et al. 1984) and interpreted as illumination of vertical magnetic
fields with mG strength (Yusef-Zadeh \& Morris 1987).

\vspace{1mm}
\noindent{\bf E. Strength of regular and random fields? There were estimates.} 

The strength of large-scale regular field is about 2~$\mu$G (HQ94,
RK89), and the total field is about 6$\mu$G. Therefore, random field
is stronger than the regular fields. % , and in the local Galactic region
%has an energy of 3.7 *** this doesn't seem correct *** times that
%of the regular fields.
By adopting a single-cell-size model for the
turbulent field, RK89 obtained a turbulent field strength of 5~$\mu$G
with a cell size of 55 pc, and Ohno \& Shibata (1993) got 4 - 6~$\mu$G
for the random fields with an assumed cell size in the range 10 - 100
pc. Heiles (1996b) discussed the strength and energy of the random
fields and regular fields.

Rand \& Lyne (1994) found evidence for stronger fields towards the 
Galactic center.

\vspace{1mm}
\noindent{\bf F. Others: Unknown.} 

There was no information about the variation of field strength with Galactocentric
radius or Galactic height, although there was some hints for such variations
(e.g. Beuermann et al. 1985). It is understandable that the fields in the
arm region could be more tangled than these in the interarm regions (see
HQ94), but not much more information is available. There was no
consideration of the spatial energy spectrum, i.e., the magnetic field
strength on different scales, although turbulence in interstellar medium was
known already.

\section{The Galactic magnetic fields: progress last decade}

Pulsars have unique advantages as probes of the large-scale Galactic
magnetic field. Their distribution throughout the Galaxy at
approximately known distances allows a true three-dimensional mapping
of the large-scale field structure. Furthermore, combined with the
measured DMs, pulsar RMs give us a direct measure of the mean
line-of-sight field strength along the path, weighted by the local
electron density. In last decade, a large number of pulsars have been
discovered by Parkes pulsar surveys (e.g., Manchester et al. 2001), and
many of them are distributed over more than half of the Galactic
disk. The RMs of these pulsars provide a unique opportunity for
investigation of the magnetic field structure in the inner
Galaxy. Compared with about 200 pulsars RMs available in 1993, now in
total, the RMs of 550 pulsars have been observed, and about 300 of
them by Qiao et al. (1995) and Han et al. (1999, 2006).

\vspace{-2mm}
\subsection{The magnetic field structure versus the spiral arms: new consensus} 
After Han \& Qiao (1994) and Indrani \& Deshpande (1998) as well as Han et
al. (1999), a bisymmetric spiral model for magnetic fields in local area
($<$ a few kpc) has been established. The new analysis of starlight
polarization data of Heiles (1996a) also gives a pitch angle of large-scale
magnetic fields about $-8^o$. We can also conclude that the large-scale
magnetic fields in our Galaxy, at least in the local region, follow the
spiral structure and probably have the same pitch angle of spiral arms.

\subsection{Discrimination of Models} 

The limited pulsar RM data and only measurements in local Galactic regions
gives room for three models to survive. Discrimination between different
models is complicated by small-scale irregularities of field structure and
sparse measurements. However, the measured pitch angle of the magnetic
fields using different approaches, as we have seen above, is hard evidence
to rule out the concentric ring model of RK89 and Rand \& Lyne (1994) which
has zero pitch angle. See detailed analysis of Indrani \& Deshpande
(1998) and discussions in Han et al. (1999, 2006). The axisymmetric model of
Vall\'ee (1996) suggests that the field near the Norma arm is clockwise, which
has been disapproved by Han et al. (2002, 2006). New measurements of the
large-scale fields in the Galactic disk (Han et al. 2006) suggest a tighter
BSS field structure.

\begin{figure}[h]
\begin{center}
\begin{minipage}{35pc}
\begin{center}
\includegraphics[angle=270,width=30pc]{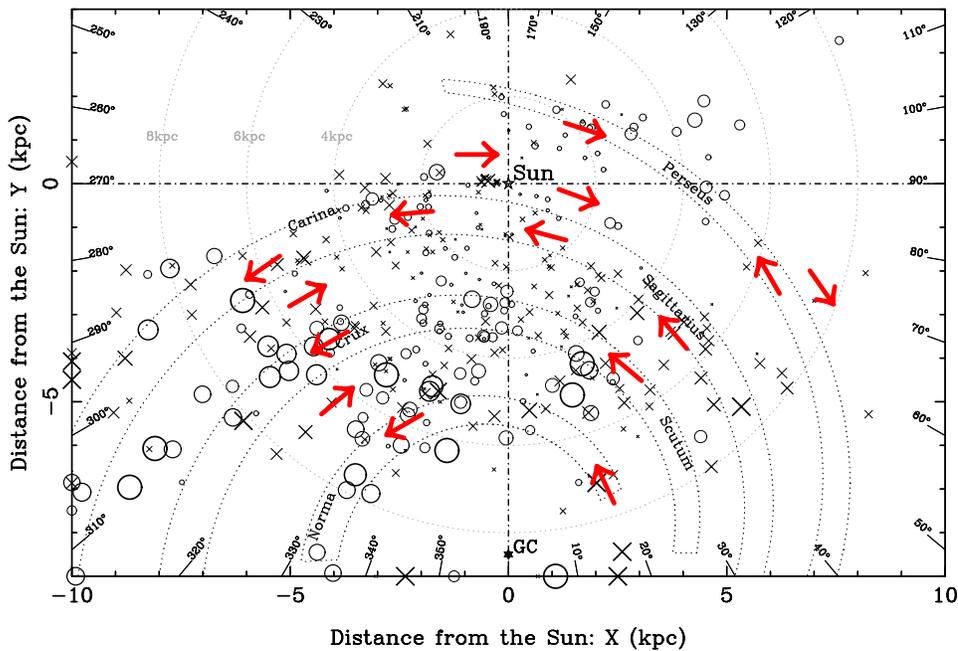}
\caption{The RM distribution of 374 pulsars with $|b|<8\degr$, projected
onto the Galactic Plane. The linear sizes of the symbols are proportional to
the square root of the RM values, with limits of 9 and 900 rad m$^{-2}$. The
crosses represent positive RMs, and the open circles represent negative
RMs. The approximate locations of four spiral arms are indicated. The
large-scale structure of magnetic fields derived from pulsar RMs are
indicated by thick arrows. See Han et al. (2006) for details.}
\end{center}
\end{minipage}
\end{center}
\vspace{-10mm}
\end{figure}
\subsection{Field structure in the Galactic disk: new measurements}

We observed more than 300 pulsar RMs (Qiao et al. 1995 and Han et al. 1999,
2006), most of which lie in the fourth and first Galactic quadrants and are
relatively distant. These new measurements enable us to investigate the
structure of the Galactic magnetic field over a much larger region than was
previously possible. We even detected counter-clockwise magnetic fields
in the most inner arm, the Norma arm (Han et al. 2002). A more complete
analysis by Han et al. (2006) gives such a picture for the coherent
large-scale fields aligned with the spiral-arm structure in the Galactic
disk, as shown in Fig.1: magnetic fields in all inner spiral arms are
counterclockwise when viewed from the North Galactic pole.  On the other
hand, at least in the local region and in the inner Galaxy in the fourth
quadrant, there is good evidence that the fields in interarm regions are
similarly coherent, but clockwise in orientation.  There are at least two or
three reversals in the inner Galaxy, occurring near the boundary of the
spiral arms (Han et al. 1999, 2006). The magnetic field in the Perseus arm
can not be determined well, though Brown et al. (2003) argued for no
reversal, using the negative RMs for distant pulsars and extragalactic
sources which in fact suggest the interarm fields both between the
Sagitarius and Perseus arms and beyond the Perseus arm are predominantly
clockwise.

\subsection{Field structure in the Galactic halo}
The magnetic field structure in halos of other galaxies is difficult to
observe. Our Galaxy is a unique case for detailed studies, since 
polarized radio sources all over the sky can be used as probes for the magnetic 
fields in the Galactic halo.
\begin{figure}[h]
\begin{center}
  \begin{minipage}[t]{0.6\linewidth}
\centerline{\psfig{figure=ers.ps,angle=270,width=80mm}}
  \end{minipage}%
  \begin{minipage}[t]{0.4\textwidth}
%\vspace{-4mm}
\centerline{\psfig{figure=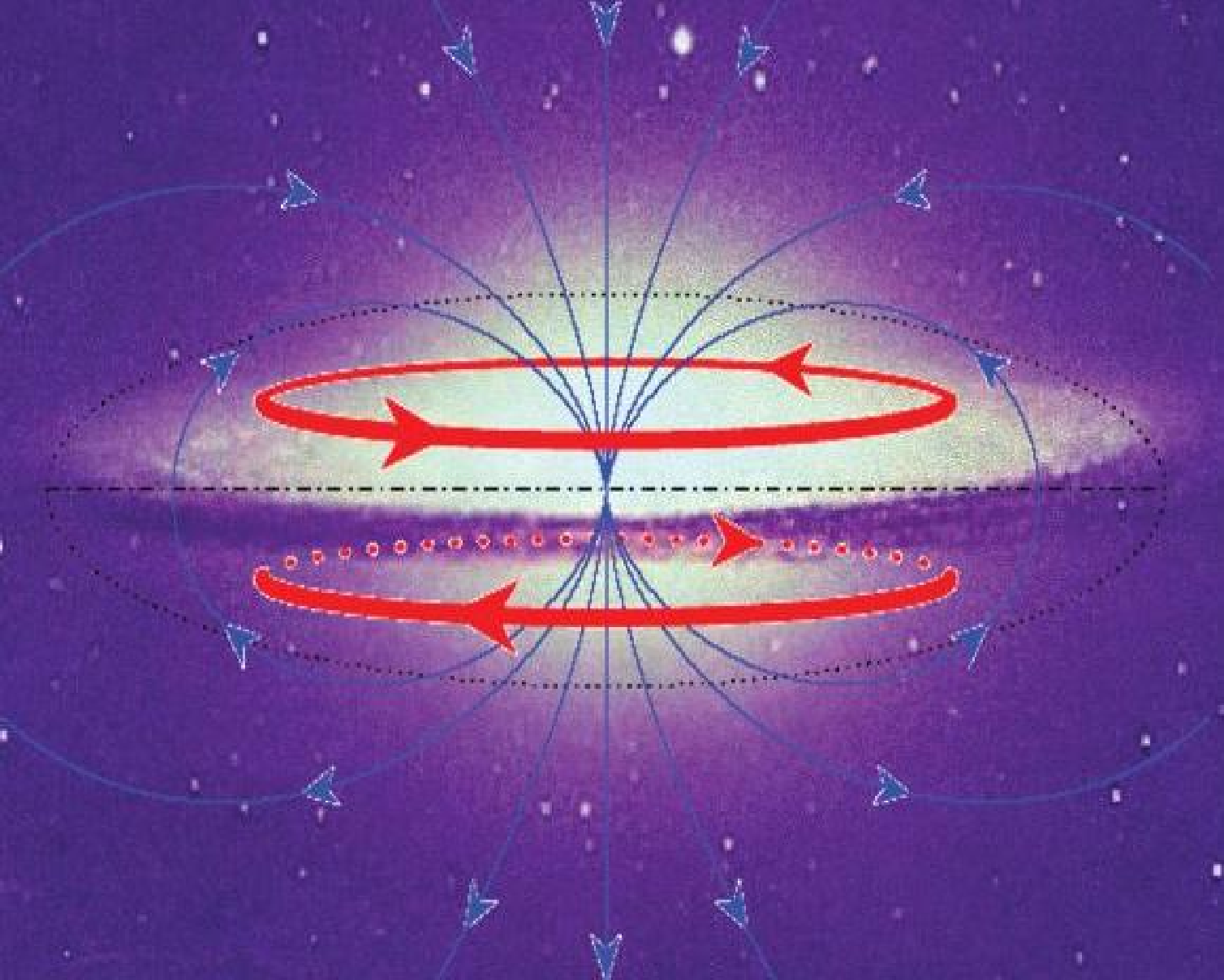,angle=0,width=50mm}}
  \end{minipage}%
\caption{The antisymmetric rotation measure sky, derived from RMs of
  extragalactic radio sources after filtering out the
  outliers of anomalous RM values, should correspond to the magnetic field 
  structure in the Galactic halo as illustrated on the right. See Han et
  al. (1997) for details.}
\end{center}
\vspace{-10mm}
\end{figure}

From the RM distribution in the sky, Han et al. (1997, 1999) identified the
striking antisymmetry in the inner Galaxy respect to the Galactic
coordinates, as being a result from the azimuth magnetic fields in the
Galactic halo with reversed field directions below and above the Galactic
plane (see Fig.2). Such a field can be naturally produced by an A0 mode of
dynamo (see Wielebinski \& Krause 1993 and Beck et al. 1996 for reviews). The observed
filaments near the Galactic center should be result from the dipole field in
this scenario.  The local vertical field component of $\sim$0.2~$\mu$G (HQ94
and Han et al. 1999) may be related to the dipole field in the solar
vicinity.

Han (2004) has shown that the RM amplitudes of extragalactic radio sources
in the mid-latitudes of the inner Galaxy are systematically larger than 
those of pulsars, indicating that the antisymmetric magnetic fields 
are not local but are extended towards the Galactic center, far beyond the pulsars.

\subsection{Field strength on different scales}
Interstellar magnetic fields exist over a broad range of spatial scales,
from the large Galactic scales to the very small dissipative
scales, but with different field strength. Knowledge of the
complete magnetic energy spectrum can offer a solid observational test for
dynamo and other theories for the origin of Galactic magnetic fields
(e.g. Balsara \& Kim 2005).

Estimation of the large-scale field strength (e.g. HQ94, Han et
al. 2006) and a turbulent field strength at a scale of tens of pc by
RK89 and Ohno \& Shibata (1993) is only the first step.  It is also
possible to get more hints from electron density fluctuations in
interstellar medium, since magnetic fields are also always frozen in
the interstellar gas. Armstrong et al. (1995) showed that the spatial
power spectrum of electron density fluctuations from small scales up
to a few pc could be approximated by a single power law with a 3D
spectral index $-$3.7, very close to the Kolmogorov spectrum.  Minter
\& Spangler (1996) found that structure functions of RM and emission
measure were consistent with a 3D-turbulence Kolmogorov spectra of
magnetic fields up to 4~pc, but with a 2D turbulence between 4~pc and
80 pc. Haverkorn et al. (2006) found the RM fluctuations are much
enhanced in the Galactic spiral arms than in interarm regions.

Pulsar RMs are the integration of field strength times electron
density over the path from a pulsar to us. Therefore, RM data of
pulsars with different distances should reflect the fluctuations on
different scales. Han et al. (2004) took RM differences and
obtained the spatial energy spectrum of the Galactic magnetic field in
scales between $0.5<\lambda<15$~kpc, which is a 1D power-law as
$E_B(k)\sim k^{-0.37\pm0.10}$, with $k=1/\lambda$. The rms field
strength is approximately 6 $\mu$G over the relevant scales and the
spectrum is much flatter than the Kolmogorov spectrum for the
interstellar electron density and magnetic energy at scales less than
a few pc (see Fig.3).

\begin{figure}[h]
  \begin{minipage}[t]{0.53\textwidth}
  \centering
\includegraphics[width=50mm,height=70mm,angle=-90]{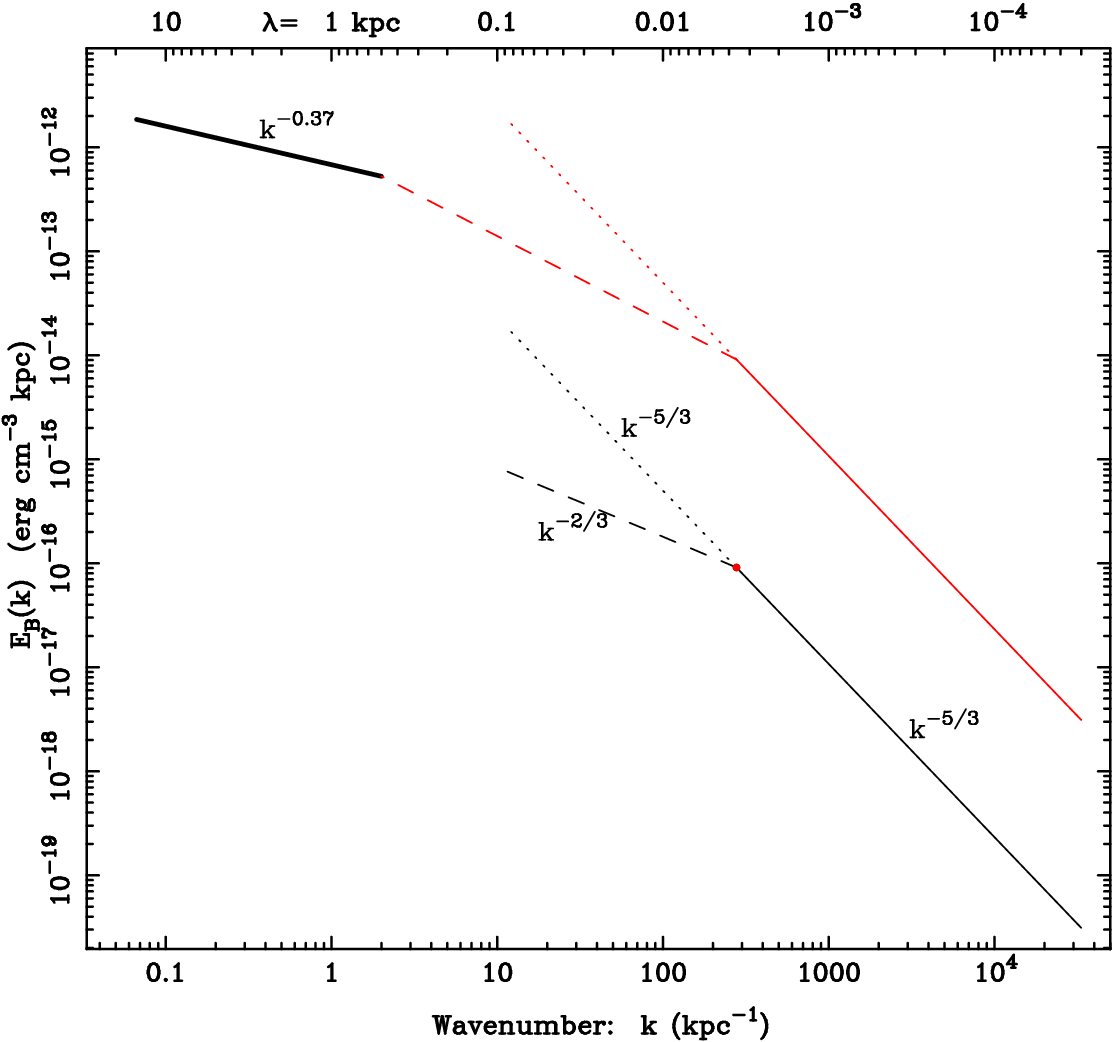}
  \vspace{-5mm}
\caption{\small Composite magnetic-energy spectrum in our Galaxy.
The large-scale spectrum was derived from pulsar RM data.
The thin solid and dashed/dotted lines at smaller scales are the
Kolmogorov and 2D-turbulence spectra inferred from the results of 
Minter \& Spangler (1996), and the upper one is from new measurements 
of Minter (2004, private email). See Han et al. (2004) for details.}
  \end{minipage}%
\hspace{0.01\textwidth}
  \begin{minipage}[t]{0.48\linewidth}
  \centering
  \vspace{2mm}
\includegraphics[angle=270,width=64mm]{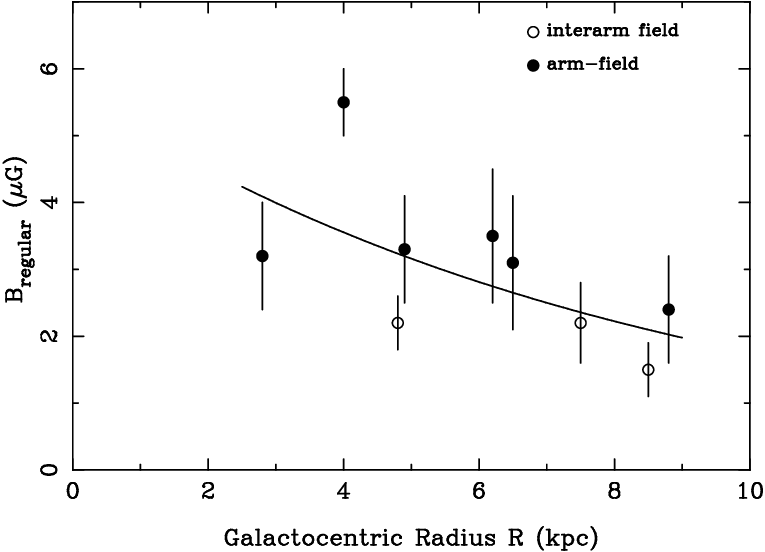}
\caption{\small Variation of the large-scale regular field strength with
the Galactocentric radius. Filled dots are for arm regions and small open
circles are for interarm regions. The curved line is a fit of an exponential
model. See Han et al. (2006) for details.}
  \end{minipage}%
\vspace{-3mm}
\end{figure}

\subsection{Variation of the field strength with Galactocentric radius}
Stronger regular magnetic fields in the Galactic disk towards the Galactic
Center have been suggested by Sofue \& Fujimoto (1983), RK89 and Heiles
(1996b). Such a radial variation of total field strength has been derived
from modeling of the Galactic synchrotron emission (E. Berkhuijsen, Fig.14
in in Wielebinski 2005) and the Galactic $\gamma-$ray background by Strong
et al. (2000). Measurements of the regular field strength in solar vicinity
give values of $1.5\pm0.4\;\mu$G (HQ94, Indrani \& Deshpande 1998), but near
the Norma arm it is $4.4\pm0.9\;\mu$G (Han et al. 2002).

%\begin{figure}[h]
%\begin{center}
%%\includegraphics[width=50mm,height=90mm,angle=270]{BR.ps}
%\includegraphics[angle=270,width=70mm]{BR.ps}
%\caption{Dependence of the strength of the large-scale regular field with
%the Galactocentric radius. Filled dots are for arm regions and small open
%circles are for interarm regions. The curved line is a fit of an exponential
%model. See Han et al. (2006) for details.}
%\end{center}
%\vspace{-5mm}
%\end{figure}

With the much more pulsar RM data now available, Han et al. (2006) were able
to measure the regular field strength near the tangential points in the 1st
and 4th Galactic quadrants, and then plot the dependence of regular field
strength on the Galactoradii (see Fig. 4). Although uncertainties are large,
there are clear tendencies for fields to be stronger at smaller
Galactocentric radii and weaker in interarm regions. To parameterize the
radial variation, an exponential function was used as following, which not
only gives the smallest $\chi^2$ value but also avoids the singularity at
$R=0$ (for $1/R$) and unphysical values at large R (for the linear
gradient). That is,
%\begin{equation}
$
B_{\rm reg}(R) =
B_0 \; \exp \left[ \frac{-(R-R_{\odot})} {R_{\rm B}} \right] ,
$
%\end{equation}
with the strength of the large-scale or regular field at the Sun,
$B_0=2.1\pm0.3$ $\mu$G and the scale radius $R_{\rm B}=8.5\pm4.7$ kpc.

\subsection{Field structure and strength near the Galactic 
center}

Progress has been made in two aspects for the region within tens to hundreds
pc of the Galactic Center, both for poloidal field and for
toroidal fields.

\noindent{\bf Poloidal fields:} More non-thermal filaments near the Galactic
center have been discovered (e.g. LaRosa et al. 2004, Nord et al. 2004,
Yusef-Zadeh 2004). The majority of the brighter non-thermal filaments are
perpendicular to the Galactic plane, indicating a predominantly poloidal
fields of $\sim$mG strength. But some filaments are not, indicating a more
complicated field structure than just the poloidal field.  LaRosa et
al.(2005) detected the diffuse radio emission and argued for a weak
pervasive field of tens of $\mu$G near the Galactic Center. The new
discovery of an infrared 'double helix' nebula by Morris et al. (2006)
reinforces the conclusion of strong magnetic fields merging from the rotated
circumnuclear gas disk near the Galactic center.

\noindent{\bf Toroidal fields:} With the development of polarimetry at
mm, submm or infrared wavelengths, toroidal fields have been observed
near the Galactic center (Novak et al. 2003, Chuss et al. 2003),
complimenting the poloidal fields shown by the vertical filaments.
Analysis of the much enhanced RMs of radio sources near the Galactic
Center (e.g. Roy et al. 2005) may indicate toroidal field structure.

\section{Concluding remarks}

In the last decade, there has been significant progress in studies of 
Galactic magnetic fields, mainly due to the availability of a large number of
newly observed RMs of pulsars. Further pulsar rotation measure observations,
especially for interarm regions and especially in the first Galactic
quadrant, would be especially valuable to confirm the large-scale magnetic
field structure in the Galactic disk. An improved RM database for the
whole sky will enable us to probe details of the magnetic fields in the
Galactic halo. Future detailed modeling of the global magnetic field 
structure of our Galaxy should match all measurements of the fields
in different directions or locations, including the field near the Galactic 
center.

%\newpage
\begin{acknowledgements}
I am very grateful to colleagues who have collaborated to make the progress
described in this review: Dr. R.N. Manchester from Australia Telescope
National Facility, CSIRO, Prof. G.J. Qiao from Peking University (China),
Prof. Andrew Lyne from Jodrell Bank Observatory (UK), and Dr. Katia
Ferri\'ere from Observatory of Midi-Pyr\'en\'ees (France).  I thank
R.N. Manchester, E. Berkhuijsen, R. Beck, Xiaohui Sun and Wolfgang
Reich for reading the draft carefully.  The author is supported by the
National Natural Science Foundation of China (10521001 and 10473015).
\end{acknowledgements}

\label{lastpage}


\small
\begin{thebibliography}{99}
%% you can type \apj for ApJ, \aap for A&A, \apss for Ap&SS, etc. Please consult
%% the macro cjaa.cls. You can also find them in aasguide.tex (AASTeX for ApJ, AJ, PASP)
%% Please follow the format of ChJAA's reference list
\vspace{-2mm}
\bibitem[{Andreasyan} and {Makarov}(1989)]{am89}
{Andreasyan}, R.~R., and {Makarov}, A.~N., 1989, Afz, 31, 247.

\vspace{-0.5mm}
\bibitem[{Armstrong {et~al.}(1995)Armstrong, Rickett, \& Spangler}]{ars95}
Armstrong, J.~W., Rickett, B.~J., \& Spangler, S.~R. 1995, ApJ, 443, 209

\vspace{-0.5mm}
\bibitem[{{Balsara} \& {Kim }(2005)}]{bk05}
{Balsara}, D. \& {Kim}, J. 2005, ApJ 634, 390

\vspace{-0.5mm}
\bibitem[\protect\citename{Beck et al. }1996]{bbm+96}
Beck R., Brandenburg A., Moss D., Shukurov A.~M.,  Sokoloff D.~D.,
1996, ARA\&A, 34, 155

\vspace{-0.5mm}
\bibitem[{{Beuermann} {et~al.}(1985)}]{bkb85}
Beuermann, K., Kanbach, G. \& Berkhuijsen, E. M. 1985, A\&A, 153, 17

\vspace{-0.5mm}
\bibitem[{{Brown} {et~al.}(2003){Brown}, {Taylor}, {Wielebinski},
  \& {Mueller}}]{btwm03}
{Brown}, J.~C., {Taylor}, A.~R., {Wielebinski}, R., \& {Mueller}, P.
  2003, ApJ, 592, L29

\vspace{-0.5mm}
\bibitem[{{Chuss et al. }(2003)}]{cdd+03}
Chuss, D. T., Davidson, J. A., Dotson, J. L. et al. 2003, ApJ, 599, 1116

%\bibitem[{{Crutcher}(2004)}]{cru04}
%{Crutcher}, R. 2004, In: The Magnetized Interstellar Medium, Copernicus GmbH, p.123

\vspace{-0.5mm}
\bibitem[{{Hamilton} \& {Lyne}(1987)}]{hl87}
{Hamilton}, P.~A. \& {Lyne}, A.~G. 1987, MNRAS, 224, 1073

\vspace{-0.5mm}
\bibitem[{{Han}(2001)}]{han01}
{Han}, J.~L. 2001, Ap\&SS, 278, 181

\vspace{-0.5mm}
\bibitem[Han (2002)]{han02}
Han, J.L. 2002, in: Astrophysical Polarized Backgrounds, AIP 609, p.98

\vspace{-0.5mm}
\bibitem[{{Han}(2004)}]{han04}
{Han}, J.~L. 2004, In: The Magnetized Interstellar Medium, Copernicus GmbH, p.3

\vspace{-0.5mm}
\bibitem[{{Han} \& {Qiao}(1993)}]{hq93}
{Han}, J.~L. \& {Qiao}, G.~J. 1993, IAU Symp.157, 279 

\vspace{-0.5mm}
\bibitem[{{Han} \& {Qiao}(1994)}]{hq94}
{Han}, J.~L. \& {Qiao}, G.~J. 1994, A\&A, 288, 759 (HQ94)

\vspace{-0.5mm}
\bibitem[{{Han} {et~al.}(2004){Han}, {Ferriere}, \& {Manchester}}]{hfm04}
{Han}, J.~L., {Ferriere}, K., \& {Manchester}, R.~N. 2004, ApJ, 610, 820

\vspace{-0.5mm}
\bibitem[{Han {et~al.}(1997)Han, Manchester, Berkhuijsen, \& Beck}]{hmbb97}
Han, J.~L., Manchester, R.~N., Berkhuijsen, E.~M., \& Beck, R. 1997, A\&A, 322,
  98

\vspace{-0.5mm}
\bibitem[{{Han} {et~al.}(1999){Han}, {Manchester}, \& {Qiao}}]{hmq99}
{Han}, J.~L., {Manchester}, R.~N., \& {Qiao}, G.~J. 1999, MNRAS, 306, 371

\vspace{-0.5mm}
\bibitem[{Han {et~al.}(2002)Han, Manchester, Lyne, \& Qiao}]{hmlq02}
Han, J.~L., Manchester, R.~N., Lyne, A.~G., \& Qiao, G.~J. 2002, ApJ, 570, L17

\vspace{-0.5mm}
\bibitem[{Han {et~al.}(2006)Han, Manchester, Lyne, Qiao, \& van Straten}]{hml+06}
Han, J.~L., Manchester, R.~N., Lyne, A.~G., Qiao, G.~J., \& van Straten,
W. 2006, ApJ, in press.

\vspace{-0.5mm}
\bibitem[{{Haverkorn} {et~al.}(2006)}]{hgb+06}
Haverkorn, M., Gaensler, B. M., Brown, J. C., et al. 2006, ApJ 637, L33

\vspace{-0.5mm}
\bibitem[{Heiles(1996a)}]{hei96}
Heiles, C. 1996a, ApJ, 462, 316

\vspace{-0.5mm}
\bibitem[{{Heiles}(1996b)}]{hei96b}
{Heiles}, C. 1996b, in ASP Conf. Ser. 97: Polarimetry of the Interstellar
  Medium, 457

\vspace{-0.5mm}
\bibitem[{{Heiles \& Crutcher} {et~al.}(2003)}]{hc05}
Heiles, C, \& Crutcher, R. 2005, In: Cosmic Magnetic Fields, LNP 664, 137

\vspace{-0.5mm}
\bibitem[{{Indrani} \& {Deshpande}(1998)}]{id98}
{Indrani}, C. \& {Deshpande}, A.~A. 1998, New Astronomy, 4, 33

\vspace{-0.5mm}
\bibitem[{LaRosa et al. (2004)}]{lnm04}
LaRosa, T. N., Nord, M. E., Lazio, T.J.W., Kassim N.E. 2004, ApJ, 607, 302

\vspace{-0.5mm}
\bibitem[{LaRosa et al. (2005)}]{lnm05}
LaRosa, T. N., Brogan C.L., Shore S.N., et al. 2005, ApJ, 626, L23

\vspace{-0.5mm}
\bibitem[{Lyne \& Smith(1989)}]{ls89}
Lyne, A.~G. \& Smith, F.~G. 1989, MNRAS, 237, 533

\vspace{-0.5mm}
\bibitem[{Manchester(1974)}]{man74}
Manchester, R.~N. 1974, ApJ, 188, 637

\vspace{-0.5mm}
\bibitem[\protect\citename{Manchester et al. }1996]{mld96}
Manchester R.~N., Lyne A.~G., D'Amico N., et al. 1996, MNRAS,
279, 1235

\vspace{-0.5mm}
\bibitem[{Manchester {et~al.}(2001)Manchester, Lyne, Camilo, Bell, Kaspi,
  D'Amico, McKay, Crawford, Stairs, Possenti, Morris, \& Sheppard}]{mlc+01}
Manchester, R.~N., Lyne, A.~G., Camilo, F., et al.  2001, MNRAS, 328, 17

\vspace{-0.5mm}
\bibitem[{Minter \& Spangler(1996)}]{ms96}
Minter, A.~H. \& Spangler, S.~R. 1996, ApJ, 458, 194

\vspace{-0.5mm}
\bibitem[{Morris et al. (2006)}]{mud06}
Morris, M., Uchida, K.,and Do, T. 2006, Nature, 440, 308

\vspace{-0.5mm}
\bibitem[{Nord} et~al.(2004)]{nlk+04}
Nord, M. E., Lazio, T. J. W., Kassim, N. E., et al. 2004, AJ, 128, 1646

\vspace{-0.5mm}
\bibitem[{Novak} et~al.(2003)]{ncr+03}
{Novak}, G., {Chuss}, D.~T., {Renbarger}, T., and {et al.} 2003, ApJ, 583, L83  

\vspace{-0.5mm}
\bibitem[{{Ohno} \& {Shibata}(1993)}]{os93}
{Ohno}, H. \& {Shibata}, S. 1993, MNRAS, 262, 953

\vspace{-0.5mm}
\bibitem[{Qiao {et~al.}(1995)Qiao, Manchester, Lyne, \& Gould}]{qmlg95}
Qiao, G.~J., Manchester, R.~N., Lyne, A.~G., \& Gould, D.~M. 1995, MNRAS, 274,
  572

\vspace{-0.5mm}
\bibitem[{Rand \& Kulkarni(1989)}]{rk89}
Rand, R.~J. \& Kulkarni, S.~R. 1989, ApJ, 343, 760 (RK89)

\vspace{-0.5mm}
\bibitem[{{Rand} \& {Lyne}(1994)}]{rl94}
{Rand}, R.~J. \& {Lyne}, A.~G. 1994, MNRAS, 268, 497

\vspace{-0.5mm}
\bibitem[{{Reich} (2006)}]{rei06}
{Reich}, W. 2006, in: Cosmic Polarization, (astro-ph/0603465), in press

%\vspace{-0.5mm}
%\bibitem[{{Reich} {et~al.}(2002){Reich}, {F{\" u}rst}, {Reich}, {Wielebinski},
%  \& {Wolleben}}]{rfr+02}
%{Reich}, W., {F{\" u}rst}, E., {Reich}, P., et al.
% 2002, in: Astrophysical Polarized Backgrounds, AIP 609, p.3

\vspace{-0.5mm}
\bibitem[{Roy et al. }(2005)]{rrs05}
Roy, S., Rao, A. P., Subrahmanyan, R. 2005, MNRAS 360, 1305

\vspace{-0.5mm}
\bibitem[{Simard-Normandin \& Kronberg(1980)}]{sk80}
Simard-Normandin, M. \& Kronberg, P.~P. 1980, ApJ, 242, 74 (SK80)

\vspace{-0.5mm}
\bibitem[{{Sofue} \& {Fujimoto}(1983)}]{sf83}
{Sofue}, Y. \& {Fujimoto}, M. 1983, ApJ, 265, 722

\vspace{-0.5mm}
\bibitem[{{Strong} {et~al.}(2000){Strong}, {Moskalenko}, \& {Reimer}}]{smr00}
{Strong}, A.~W., {Moskalenko}, I.~V., \& {Reimer}, O. 2000, ApJ, 537, 763

\vspace{-0.5mm}
\bibitem[{{Thomson} \& {Nelson}(1980)}]{tn80}
{Thomson}, R.~C. \& {Nelson}, A.~H. 1980, MNRAS, 191, 863

\vspace{-0.5mm}
\bibitem[{{Vall{\'e}e}(1996)}]{val96}
{Vall{\'e}e}, J.~P. 1996, A\&A, 308, 433

\vspace{-0.5mm}
\bibitem[{{Wielebinski }(2005)}]{wie05}
{Wielebinski}, R. 2005, in: Cosmic Magnetic Fields, LNP 664, 89

\vspace{-0.5mm}
\bibitem[\protect\citename{Wielebinski \& Krause }1993]{wk93}
Wielebinski R., Krause F., 1993, A\&AR, 4, 449

\vspace{-0.5mm}
\bibitem[{Yusef-Zadeh} and {Morris}(1987)]{ym87a}
{Yusef-Zadeh}, F., and {Morris}, M. 1987, ApJ, 320, 545.

\vspace{-0.5mm}
\bibitem[{Yusef-Zadeh} et~al.(1984)]{ymc84}
{Yusef-Zadeh}, F., {Morris}, M., and {Chance}, D. 1984, Nature, 310,
  557.

\vspace{-0.5mm}
\bibitem[{Yusef-Zadeh} et~al.(2004)]{yhc04}
{Yusef-Zadeh}, F., {Hewitt}, J.M., and {Cotton}, W. 2004, ApJS, 155, 421.

\end{thebibliography}
\end{document}